\documentstyle[12pt]{article}
\input psfig
\def\be{\begin{equation}}
\def\ee{\end{equation}}          
\def\ba{\begin{array}} 
\def\ea{\end{array}}
\def\beqn{\begin{eqnarray}}
\def\eeqn{\end{eqnarray}}
\def\bt{\begin{tabular}}
\def\et{\end{tabular}}
\def\bc{\begin{center}}
\def\ec{\end{center}}

\def\vud{$|V_{ud}|$}
\def\vus{$|V_{us}|$}
\def\vcb{$|V_{cb}|$} 

\def\vcd{$|V_{cd}|$}
\def\vcs{$|V_{cs}|$}

\def\rub{$|\frac {V_{ub}}{V_{cb}}|$}
\def\mu{$m_u$}

\def\sin2{sin$2\beta$}
\def\b{$\beta$}
\def\del{$\delta$}

\begin{document}
\title{Constructing ``Reference'' Triangle through Unitarity 
of CKM Matrix}
\author{Monika Randhawa and Manmohan Gupta \\
{\it Department of Physics,}\\
{\it Centre of Advanced Study in Physics,}\\
 {\it Panjab University, Chandigarh-
  160 014, India.}}
  \maketitle
\begin{abstract}
Motivated by the possibility of the low value of \sin2~ in the 
measurements of BABAR and BELLE collaborations, 
 a reference unitarity triangle is constructed  
using the  unitarity of the CKM matrix and the
experimental values of the well known CKM elements,
without involving any  inputs from the processes which might include  
the new physics effects. The angles of the triangle are evaluated by finding
the CP violating phase $\delta$ through the Jarlskog's rephasing
invariant parameter $J$. The present data and the unitarity of the
CKM matrix gives for $\delta$ the range  28$^o$ to 152$^o$,
which for \sin2~ translates to the  
 range 0.21 to 0.88. This range is
broadly in agreement with the recent BABAR and BELLE results. However,
a value of \sin2$\leq$0.2, advocated by Silva and Wolfenstein
 as a benchmark for new physics, would 
imply a violation in the three generation unitarity and would hint
towards the existence of a fourth generation. Further, the future 
refinements in the CKM elements will push the lower limit on \sin2~
still higher.
\end{abstract}
The recent measurements of the time dependent CP asymmetry 
$a_{\psi K_S}$ in $B^o_d({\bar B}^o_d) \rightarrow \psi K_S$
 decay by BABAR and BELLE collaborations, for example,
 \beqn a_{\psi K_S}& =&0.12 \pm 0.37 \pm 0.09~~~~~~~~  {\rm BABAR}~ 
\cite{babar},  \label{babar} \\
 a_{\psi K_S}&=&0.45^{+0.43~+0.07}_{-0.44~-0.09}~~~~~~~~~~~~~~ {\rm BELLE}~ 
\cite{belle}, \label{belle} \eeqn
 look to be smaller compared to the CDF measurements \cite{cdf}, 
for example,
\be a_{\psi K_S}^{{\rm CDF}}=0.79^{+0.41}_{-0.44}~, \label{cdf} \ee
as well as compared to the recent standard analysis of the unitarity 
triangle \cite{burasrev} with  
$|\epsilon_K|$, \rub, $\Delta m_d$ and $\Delta m_s$
as input, given as
\be a_{\psi K_S}^{{\rm SM}} = 0.67 \pm 0.17.
\label{burasrev} \ee
In the Standard Model, $a_{\psi K_S}$  is related to the angle $\beta$
of the unitarity triangle as,
\be  a_{\psi K_S} = Sin2\beta.  \label{sin2} \ee

Recently,  several authors \cite{kagan} - \cite{nxb}  have
 explored the implications of the possibility of low
value of \sin2~ in comparison to the CDF measurements 
as well as to the global analysis of the unitarity  triangle.
These analyses lead to the general consensus that the
possibility of new physics could be more prominent 
in the loop dominated  processes, in particular the 
$B_o - \bar{B_o}$ mixing. Further, it is realized that the new 
physics will not affect the tree level decay processes and 
the unitarity of the three generation CKM matrix in the SM approaches
as well as in its extensions \cite{kagan}-\cite{ut5}. 
In this connection, for better appraisal of new physics,
it has been generally recommended  to construct a universal 
or reference unitarity triangle \cite{kagan},\cite{ut1}-\cite{ut5},
wherein  the inputs are free from the processes which might include  
the new physics effects, in particular the
$B_o - \bar{B_o}$ mixing and $K_o - \bar{K_o}$ mixing parameters.
Keeping this in mind several strategies, 
model dependent \cite{ut1,ut2} as well as model independent 
\cite{ut3,ut4,ut5}, have been formulated to construct the triangle,
 however by and large both approaches rely on the 
rare decays. The reference triangle to be constructed is defined as,
\be V_{ud}V_{ub}^{*} + V_{cd}V_{cb}^{*} + V_{td}V_{tb}^{*} = 0,
\label{db} \ee
 obtained by employing the orthogonality of the
 first and third column of the CKM matrix (henceforth referred to as
 triangle $db$).
In this triangle the elements involving $t$ quark
have not been experimentally measured as yet and hence to 
construct the   triangle, the inputs from rare decays
involving elements $V_{td}$ and $V_{tb}$ through loops 
have to be used.

In this context, it is interesting to note that
despite several analyses of the CKM
phenomenology in the past \cite{burasrev}, \cite{jarlskog}
-\cite{parodi}
 yielding valuable information, 
the implications of three generation unitarity have not been
examined in detail  in the construction of the reference triangle.
A reference triangle constructed purely from the considerations of
unitarity as well as using experimentally measured CKM elements
 will be free from the effects of new physics and hence could serve
as a tool for deciphering deviation from the SM in measuring the
CP asymmetries.

 The purpose of the present communication is to construct
the  triangle $db$ using unitarity of the three generation CKM matrix
 by evaluating the   Jarlskog's Rephasing Invariant Parameter $J$ 
and consequently the  CP violating phase $\delta$. In particular,
we intend to evaluate angles $\alpha$, $\beta$ and $\gamma$
of the triangle $db$ and study the implications of the low value of \sin2~
for unitarity.

   To begin with we consider the six non diagonal relations
     implied by the unitarity of the
CKM matrix. One of the relations is mentioned above in equation
\ref{db} and the other five are as follows,
\beqn
  ds~~~~~~~V_{ud}V_{us}^{*} + V_{cd}V_{cs}^{*} + V_{td}V_{ts}^{*} = 0, 
\label{ds}  \\
  sb~~~~~~~V_{us}V_{ub}^{*} + V_{cs}V_{cb}^{*} + V_{ts}V_{tb}^{*} = 0,
\label{sb}  \\
  ut~~~~~~~V_{ud}V_{td}^{*} + V_{us}V_{ts}^{*} + V_{ub}V_{tb}^{*} = 0,  
\label{ut}  \\
  uc~~~~~~~V_{ud}V_{cd}^{*} + V_{us}V_{cs}^{*} + V_{ub}V_{cb}^{*} = 0,  
\label{uc}  \\
  ct~~~~~~~V_{cd}V_{td}^{*} + V_{cs}V_{ts}^{*} + V_{cb}V_{tb}^{*} = 0.
\label{ct}
\eeqn
 The letters before the different equations denote the respective
 triangles.

\begin{figure}
   \centerline{\psfig{figure=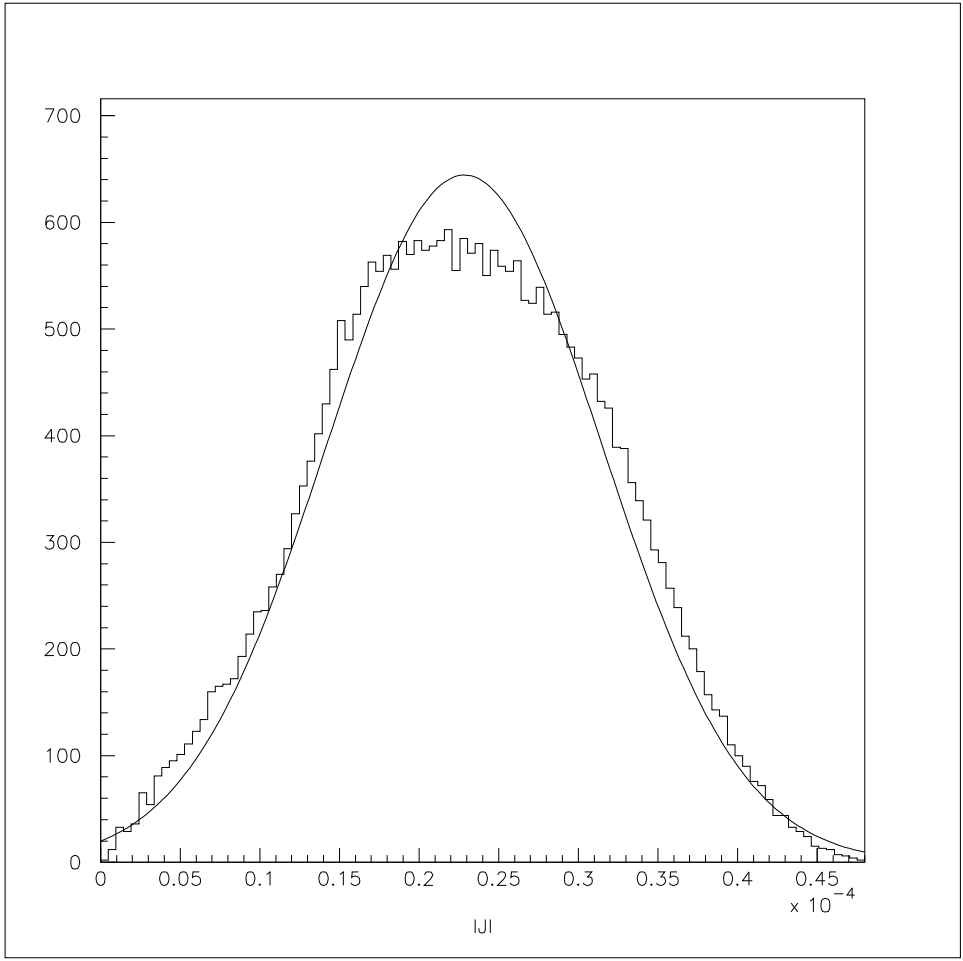,width=5in,height=5in}}
   \caption{Gaussian fitted  to the histogram of $|J|$ generated
   by considering the triangle $uc$ with the input constraints
 $|a|+|c| > |b|$ and $|b|+|c| > |a|$, where $a = |V_{ud}^*V_{cd}|$,
  $b=|V_{us}^*V_{cs}|$ and 
  $c=|V_{ub}^*V_{cb}|$.}
  \label{fig1}
  \end{figure}

As mentioned above, in the triangle $db$ the elements
 $V_{td}$ and $V_{tb}$ are not  experimentally
measured, therefore to obtain these elements without involving inputs
from $K_o - \bar{K_o}$  and  $B_o - \bar{B_o}$ mixing and rare decays one
needs to make use of the PDG \cite{pdg} representation of the
CKM matrix given below,
  \be V_{CKM}= \left( \ba {lll} c_{12} c_{13} & s_{12} c_{13} &
  s_{13}e^{-i\delta} \\
  -s_{12} c_{23} - c_{12} s_{23} s_{13}e^{i\delta} &
 c_{12} c_{23} - s_{12} s_{23}s_{13}e^{i\delta}
  & s_{23} c_{13} \\
  s_{12} s_{23} - c_{12} c_{23} s_{13}e^{i\delta} &
  - c_{12} s_{23} - s_{12}c_{23} s_{13}e^{i\delta} &
  c_{23} c_{13} \ea \right),  \label{ckm} \ee
  with $c_{ij}=cos\theta_{ij}$ and   $s_{ij}=sin\theta_{ij}$ for 
 $i,j=1,2,3.$
 Since one can obtain $s_{12}$, $s_{23}$ and $s_{13}$
  from the experimentally well known elements
\vus, \vcb~  and
 \rub~  given in Table
 \ref{tabinput}, the CP violating phase
 $\delta$ remains the only unknown parameter in determining the triangle
 $db$, which  is related to the  Jarlskog's rephasing invariant
 parameter $J$ as,
      \be J = s_{12}s_{23}s_{13}c_{12}
        c_{23}c^2_{13}sin \delta. \label{j} \ee
An evaluation of $J$ would allow us to find $\delta$ and 
consequently the angles  $\alpha$, \b~ and $\gamma$ of the 
 triangle $db$. To evaluate $J$, we make use of the fact 
that the areas of all the six triangles (equations \ref{db}-\ref{ct})
are  equal and that the area of any of the unitarity triangle
 is related to Jarlskog's Rephasing Invariant Parameter $J$ as,
\be J = 2 \times {\rm Area~ of~any~ of~ the~ Unitarity~ Triangle.}
 \label{area} \ee
This, therefore affords an opportunity to evaluate $J$
through one of the unitarity triangle whose sides are 
experimentally well known, for example, triangle $uc$.  
  The triangle $uc$ though is quite well known, but it is highly 
squashed, therefore one needs to be careful while
evaluating $J$ through this triangle.
The sides of the triangle  represented by $|V_{ud}^*V_{cd}|~(=a)$ and
  $|V_{us}^*V_{cs}| ~(=b)$ are of comparable lengths while the third side
  $|V_{ub}^*V_{cb}| ~(=c)$  is several orders of
    magnitude smaller compared to $a$ and $b$.
  This creates complications for evaluating the area of the triangle
  without violating the existence of CP violation.
  These complications can be avoided without violating the
unitarity by incorporating the constraints
   $|a|+|c| > |b|$ and $|b|+|c| > |a|$ \cite{branco}.
    Using these  constraints and the experimental data given in the
     table \ref{tabinput}, a histogram can be generated, 
shown in figure \ref{fig1}, to which a gaussian is 
fitted yielding the result,
     \be |J|= (2.59 \pm 0.79) \times 10^{-5}
          \label{jpdg1s}.  \ee
  This value of $|J|$
  can now be used to calculate $\delta$ using the
   equation \ref{j}, which can be re-written as,
      \be J = J^{'} sin \delta  \label{jpdg}, \ee where,
        \be J^{'} = s_{12}s_{23}s_{13}c_{12}
        c_{23}c^2_{13}. \label{j'} \ee
  Calculating $s_{12},~ s_{23}$ and $s_{13}$ from
    the experimental values of  \vus, \rub,~
    and  \vcb~ given in table \ref{tabinput} and
   following the procedure outlined above for evaluating $|J|$,
     $J^{'}$ comes out to be,
      \be J^{'}= (3.23 \pm 0.63) \times 10^{-5}.
           \label{xpdg1s} \ee
Since $J^{'}sin \delta$ should reproduce  $|J|$
 calculated through the unitarity triangle $uc$, therefore  
 comparing equations \ref{jpdg1s} and \ref{xpdg1s}, one can easily find 
    out the widest limits on $\delta$, for example,
 \be \delta = 28^o ~{\rm to}  ~152^o. \label{deluni} \ee
       This value of $\delta$ apparently looks to be the consequence
  only of the unitarity relationship given by equation \ref{uc}.
    However on further investigation, as shown by Branco and Lavoura
  \cite{branco}, one finds that this $\delta$ range is consequence
   of all the non trivial unitarity constraints. In this sense the above
  range could be attributed to as a consequence of unitarity of the
 CKM matrix. It needs to be noted that with the above range of $\delta$
 and  the  experimental  values of \vus, \vcb~ and  
\rub~ given in Table \ref{tabinput}, the CKM matrix thus evaluated
 is in excellent agreement with  PDG  CKM matrix \cite{pdg}.
    
Alternatively, using equation \ref{jpdg}, one can plot a
   histogram for $\delta$ as well, to which fitting a Gaussian yields,
 \beqn \delta & = & 50^o \pm 20^o~ ({\rm I~ quadrant}), \nonumber \\
     & &  130^o \pm 20^o~ ({\rm II~ quadrant}). \label{dhis68} \eeqn
 This gives us relatively stronger bounds on $\delta$.  
 However, to be conservative,  we have used the range of $\delta$
 as given by  equation \ref{deluni} for the
 subsequent calculations.
 
 After having obtained a range for $\delta$, the triangle $db$
 can be constructed, however  without involving inputs from
 the phenomena which may have influence from the new physics as
 well as without the inputs from the rare decays.
The angles $\alpha$, \b~ and $\gamma$ of the triangle can be
expressed  in terms of the CKM elements as,  
\be \alpha = arg\left(\frac{-V_{td}V_{tb}^*}{V_{ud}V_{ub}^*} \right), 
\label{alpha} \ee
\be \beta = arg\left(\frac{-V_{cd}V_{cb}^*}{V_{td}V_{tb}^*}\right),
 \label{beta} \ee
\be \gamma = arg\left(\frac{-V_{ud}V_{ub}^*}{V_{cd}V_{cb}^*}\right),
 \label{gamma} \ee
where CKM elements are as given by the PDG representation in the
equation \ref{ckm}. In the Table \ref{tabinput} we 
have listed the experimental values of the
CKM elements as given by PDG \cite{pdg} as well as their future values.
 Making use of the PDG representation of CKM matrix given in
equation \ref{ckm}, experimental values of  \vus,~ \vcb~
and \rub~ from table \ref{tabinput} and the range of
\del ~ given by equation \ref{deluni}, one can easily find out 
the corresponding ranges for the three angles.
In the Table \ref{tab1}, we have listed the corresponding results for 
$J$, $\delta$,  $\alpha$, \b~ and $\gamma$.
The  ranges for $\alpha$, \b~ and $\gamma$ are as follows,
\beqn \alpha \simeq 19^o~ {\rm to}~ 142^o \label{alphauni}, \\
 \beta \simeq 6^o ~{\rm to} ~ 31^o \label{betauni}, \\
 \gamma \simeq 28^o ~{\rm to} ~ 152^o. \label{gammauni} \eeqn
 While evaluating the three angles, we have taken care that the triangle
is closed.
The range of  \sin2~ corresponding to equation \ref{betauni} is given as,
 \be sin2\beta =  0.21~ {\rm to} ~0.88 \label{sin2uni}. \ee
It needs to be emphasized that this range for \sin2~ is obtained by 
making use of unitarity and the well known CKM elements listed
 in Table \ref{tabinput}. The above range has considerable overlap
with the BABAR and BELLE results, 
however if \sin2~ is found to be $\leq$0.2, a benchmark for new physics
as advocated by Silva and Wolfenstein \cite{silva}, then one may 
conclude that even the three generation unitarity may not be valid and
one may have to go to four generations to explain the low values of \sin2.
In such a scenario, the widely advocated assumption \cite{kagan}
-\cite{ut5} that the non SM physics
resides in loop dominated processes only may not be valid.
 
 A few comments are in order.
It is interesting to examine the consequences of the future refinements 
in  the CKM elements. 
While listing the future values of the elements we have considered only
 those elements where the present error is more than 15$\%$,
for example \rub~ and \vcs. The future values of these
elements are listed in column III of Table 
\ref{tabinput}.
One finds from the Table \ref{tab1} that the refinements in  \rub~ 
and \vcs~   would improve the lower bound on \sin2~ from 0.21 to 0.31.
This would give a clear signal for physics beyond the SM 
in case \sin2~ is measured to be $\leq$ 0.2. To emphasize this conclusion,
we have also considered all the future inputs at their 90$\%$ CL 
and this gives the lower limit of \sin2=0.18.

It may be of interest to mention that a recent investigations
involving texture 4 zeros quark mass matrices 
and unitarity \cite{massmat}, 
yield the following range for \sin2, 
\be Sin2\beta = 0.27~ {\rm to}~ 0.60, \label{massmat} \ee
which looks to be compatible with the present unitarity based 
calculations. A value of \sin2 $\leq$ 0.2 therefore, will have far
 reaching consequences for unitarity as well as for texture
 specific mass matrices \cite{massmat1}.

It is interesting to compare our results (equation \ref{sin2uni}) with
those of Buras  (equation \ref{burasrev}), obtained 
from the measurements of $|\epsilon_K|$, \rub, $\Delta m_d$ 
and $\Delta m_s$, which look to be much
narrower compared to ours. This is easy to understand when one considers 
the definition of \b~ given in equation \ref{beta}, wherein the magnitude 
and phase of $V_{td}$ play an important role. For example, the 
range of $\delta$ given by equation \ref{deluni} yields the
$V_{td}$ range as 0.0045 to 0.0135, whereas the 
range corresponding to Buras's analysis is 0.0067 to 0.0093,
which is narrower
primarily due to restrictions imposed by $|\epsilon_K|$, $\Delta m_d$ 
and $\Delta m_s$.

To conclude, we have constructed a reference unitarity triangle by making 
use of the three generation unitarity of the CKM matrix and the
experimental values of the well known CKM elements,
without involving any  inputs from the processes which might include  
the new physics effects, in particular the
$B_o - \bar{B_o}$ mixing and $K_o - \bar{K_o}$ mixing parameters
as well as the rare decays. The angles of the triangle have been
 evaluated by finding the CP violating phase $\delta$ through the 
Jarlskog's rephasing invariant parameter $J$.
The range of $\delta$ comes out to be 28$^o$ to 152$^o$ and the
corresponding range for \sin2~ is 0.21 to 0.88. This range is
broadly in agreement with the recent BABAR and BELLE results and
also has considerable overlap with the range found from the
texture 4 zeros quark mass matrices and the unitarity of the 
CKM matrix. However,
a value of \sin2$\leq$0.2 advocated by Silva and Wolfenstein
 as a benchmark for new physics would 
imply a violation in the three generation unitarity and would hint
towards the existence of a fourth generation. Further, the future 
refinements in the CKM elements will push the lower limit on \sin2~
still higher, for example from 0.21 to 0.31, thus
 giving a clear signal for physics beyond the SM 
in case \sin2~ is measured to be $\leq$ 0.2. This remains valid
even when the future values are considered at their 90$\%$ CL.
 \vskip 1cm
  {\bf ACKNOWLEDGMENTS}\\

M.G. would like to thank S.D. Sharma for useful discussions.
M.R. would like to thank CSIR, Govt. of India, for
 financial support and also the Chairman, Department of Physics,
for providing facilities to work in the department.

\newpage

\begin{table} 
\bc \begin{tabular}{|l|l|l|} \hline
Parameter & PDG values \cite{pdg} & Future values  \\ \hline
 \vud & 0.9735 $\pm$ 0.0008 & 0.9735 $\pm$ 0.0008 \\ 
\vus &  0.2196 $\pm$ 0.0023 & 0.2196 $\pm$ 0.0023 \\
\vcd & 0.224 $\pm$ 0.016 &  0.224 $\pm$ 0.016 \\
 \vcs & 1.04 $\pm$ 0.16 & 1.04 $\pm$ 0.08 \\
 \vcb & 0.0402 $\pm$0.0019 &  0.0402 $\pm$0.0019\\
 \rub &  0.090 $\pm$ 0.025 & 0.090 $\pm$ 0.010 \\ 
 & & \\ \hline
\end{tabular}
\caption{Values of the CKM parameters used throughout the paper.}
\label{tabinput}
\ec \end{table}

\begin{table}
\bc
\begin{tabular}{|c|c|c|c|} \hline
&  With PDG values  & 
 With future values &  \bt{c} With future values \\ at their
 90$\%$ CL \\ \et \\ \hline
& & & \\
 $J$ & $ (2.59 \pm 0.79) \times 10^{-5} $ &
$(2.79 \pm 0.49) \times 10^{-5} $ & $ (2.61 \pm 0.78)
 \times 10^{-5}$ \\ & & & \\
 $\delta$ & $28^o$ to $152^o$ & 
 $42^o$ to $138^o$  & $30^o$ to $150^o$ \\ &  & & \\
$\alpha$ & $19^o$ to $141^o$ &
$28^o$ to $124^o$ & $19^o$ to $143^o$ \\ & & & \\
 $\beta$ & $ 6^o$ to $31^o$  & 
 $9^o$ to $31^o$ & $5^o$ to $36^o$ \\ & & & \\
$\gamma$ &  $28^o$ to $152^o$ &
$42^o$ to $138^o$  & $30^o$ to $150^o$ \\ \hline
\end{tabular} 
\caption{$J$, $\delta$ and corresponding $\alpha$, $\beta$
and $\gamma$ with PDG and the future values of 
input parameters listed in table \ref{tabinput}}
\label{tab1}
\ec \end{table}

\newpage 

\end{document}